\documentclass[prd, aps, nofootinbib,preprintnumbers, showpacs, superscriptaddress]{revtex4}

\usepackage{latexsym}
\usepackage{amssymb}
\usepackage{amsfonts}
\usepackage{amsmath}
\usepackage{bm}
\usepackage[dvips]{graphicx}
\usepackage{color}
\usepackage{times}

\newcommand{\sgn}{\operatorname{sgn}}

\allowdisplaybreaks

\begin{document}

\title{Kaluza-Klein Cosmology: the bulk metric}

\date{\today}
\label{firstpage}

\author{Carles Bona}
\affiliation{
Universitat Illes Balears, Institute for Computational Applications with Community Code (IAC3), Palma 07122, Spain
}
\affiliation{
 Institute for Space Studies of Catalonia, Barcelona 08034, Spain
}

\author{Miguel Bezares}
\affiliation{
Universitat Illes Balears, Institute for Computational Applications with Community Code (IAC3), Palma 07122, Spain
}

\begin{abstract}
The Cosmological Principle is applied to a five-dimensional vacuum manifold. The general (non-trivial) solution is explicitly given. The result is a unique metric, parametrized with the sign of the space curvature ($k=0,\pm 1$) and the signature of the fifth coordinate. Friedmann-Robertson-Walker (FRW) metrics can be obtained from this single 'mother' metric (M-metric), by projecting onto different space-homogeneous four-dimensional hypersurfaces. The expansion factor $R$ is used as time coordinate in order to get full control of the equation of state of the resulting projection. The embedding of a generic (equilibrium) mixture of matter, radiation and cosmological constant is given, modulo a quadrature, although some signature-dependent restrictions must be accounted for.
In the 4+1 case, where the extra coordinate is spacelike, the condition ensuring that the projected hypersurface is of Lorentzian type is explicitly given. An example showing a smooth transition from an Euclidian to a Lorentzian 4D metric is provided. This dynamical signature change can be considered a classical counterpart of the Hartle-Hawking 'no-boundary' proposal. The resulting FRW model shows an initial singularity at a finite value of the expansion factor $R$. It can be termed as a 'Big unfreeze', as it is produced just by the beginning of time, without affecting space geometry. The model can be extended in order to fit the present value of the density parameters.
\end{abstract}

\maketitle


\section{Introduction}

Since the pioneering work of Kaluza-Klein, adding extra dimensions to the standard (4D) spacetime has shown to be a good strategy in the quest for unification. In modern particle physics, this has led to the brane-world models~\cite{Arkani, Randall-Sundrum, Maartens_living}, where the basic physics scenario is a higher-dimensional 'bulk' spacetime, in which gravity acts, although ordinary matter and fields are supposed to be confined in 4-dimensional 'branes' (see ref.~\cite{Carter} for a purely classical description). This approach evokes Plato's Cavern allegory, because the observed physical fields are seen just as projections (shadows) on the branes (the cavern walls), whereas the real objects are out of sight, living in some higher-dimensional reality (the bulk). The unifying power comes mainly from the fact that a single object can project different shadows in different walls. That is, different physics in the brane can be obtained just by considering different projections of a single bulk object.

In our case, this object is the 5D manifold geometry and the shadows depict the matter-energy content of the 4D spacetime. This has been termed as spacetime-matter (STM) approach (see~\cite{Overduin} for a review). It has also been implemented in Cosmology, after the pioneering work of Ponce de Leon~\cite{JPL88} (see~\cite{Wesson_book} for a review). In some of these works, different projections of the same bulk metric are considered~\cite{JPL, Camera}, so that different FRW metrics are obtained just by projecting onto different branes. A further step in that direction has been taken recently~\cite{3+2 paper}: when embedding spatially flat FRW metrics in a bulk with 3+2 signature (timelike extra coordinate), a single bulk metric was obtained, termed as 'mother metric' or rather M-metric. The aim of this paper is to generalize this unifying result to all types of spatial curvature ($k=0,\pm 1$) and bulk signature (either spacelike or timelike extra coordinate).

We will adopt a top-down approach, starting from the natural generalization of the Cosmological Principle to the five-dimensional bulk manifold. We will assume both isotropy and homogeneity with respect to three space coordinates (the ones that can be observed on the brane). Moreover, we will assume that the cosmological scale factor $R$ is a valid time coordinate. In the 3+2 case, this just amounts to a suitable coordinate choice in the time plane. In the 4+1 case, this amounts to require that the constant $R$ hypersurfaces be space-like. The cosmological expansion will affect then to all physical (timelike) observers.

With this only hypothesis, we write down in section~2 the general form of the 5D line element and the corresponding field equations in the evolution formalism. The general solution is obtained in section~3; it happens to be a single metric, which we call 'M-metric' because it is the generalization of the one previously obtained~\cite{3+2 paper} in the 3+2 case for $k=0$. Our result actually applies to all curvature cases ($k=0,\pm 1$) and for both spacelike and timelike extra dimensions. We show that this metric has a Killing vector, whose integral lines are orthogonal to the $R$ coordinate (time) lines. This means that the coordinates in which we explicitly write down the M-metric have a clear geometrical and physical meaning.

We study in section~4 different brane projections of the M-metric, leading to different FRW models. Apart from the more conventional models (open universes starting from a Big Bang), we can generate in this way some other (more exotic) cases. In the 3+2 case, we can get regular universe models, where the Big Bang lies in the infinite past, or even emergent models, starting from a quasi-stationary state~\cite{Ellis_Maartens, Ellis}. In the 4+1 case, we can generate smooth signature transitions from Euclidean to Lorentzian metrics in the brane~\cite{Ellis_signature,Mars}, which can be considered as the classical counterpart of the Hartel-Hawking 'no boundary' proposal~\cite{Hartle_Hawking}. A simple example model is presented, in the spatially flat case, in order to visualize all the details in a explicit way.

Finally, in section~5, we point out that using the expansion factor as a time coordinate provides a direct control of the matter-energy content in the brane, allowing to impose suitable equations of state. We study the case of an equilibrium combination of matter, radiation and cosmological constant. The explicit expression of the bulk metric embedding is provided, modulo a quadrature. In the process, we identify some signature restrictions: in the 3+2 case, one can get this generic mixture only in the closed universe case ($k=+1$). This confirms some previous results~\cite{3+2 paper}, in the sense that Campbell's theorem~\cite{Campbell} depends on some assumptions that can not be overlooked.


\section{5D Cosmological framework: evolution formalism}

We will consider here 5-dimensional (5D) vacuum metrics, where the extra coordinate is labeled by $\psi$. In our case, where we assume homogeneity and isotropy of the physical 3-space, this 'bulk' metric would read
\begin{equation}\label{bulk metric}
    ds^2 = \epsilon\, A^2(\psi,t)\, d\psi^2 - N^2(\psi,t)\, dt^2 + R^2(t)\;\gamma_{ij}\; dx^i\,dx^j,
\end{equation}
where the three-dimensional metric $\gamma_{ij}$ is of constant curvature, that is
\begin{equation}
    ^{(3)}\!R_{ij} = 2k\;\gamma_{ij}\;\;\;\;\;(k\,=\,0,\pm 1),
\end{equation}
and $\epsilon=\pm 1$, so that the $\psi$ coordinate can be either space-like (4+1 case, $\epsilon = 1$) or time-like (3+2 case, $\epsilon = -1$).

Note that we have chosen the $t$-coordinate lines orthogonal to the constant $R$ hypersurfaces. This can always be done in the 3+2 case, where it would be just a coordinate choice in the $(\psi,t)$ time plane. But the 4+1 case requires assuming that the resulting coordinate lines are actually timelike. This amounts to discard the possibility of having a system of physical observers moving along constant $R$ hypersurfaces.

We will consider a slicing of the 5D manifold by the family of constant $\psi$ hypersurfaces (evolution formalism). On every slice we will recover a FRW line element, namely
\begin{equation}\label{FRW}
    - N^2(\psi,t) \,dt^2 + R^2(t)\;\gamma_{ij}\; dx^i\,dx^j \equiv g_{ab}\; dx^a\,dx^b\,,
\end{equation}
where $a,b\,=\, 1,2,3,4$.
The corresponding extrinsic curvature $K_{ab}$  can be easily calculated in our case:
\begin{equation}\label{extrinsic curvature}
    K_{ab}\equiv - \frac{1}{2A} \partial_\psi ~ g_{ab} = -\frac{N'}{AN}\, u_a\,u_b\,,
\end{equation}
where the primes stand for $\psi$ derivatives and $u^a$ is the future-pointing time unit vector (the FRW metric four-velocity)
\begin{equation}\label{fourvelocity}
    u^a\,=\, \frac{1}{N}\; \delta^a_{(t)}\,,
\end{equation}
which of course verifies
\begin{equation}\label{uderivatives}
    \nabla_a\,u_b\,=\frac{\dot{R}}{NR}\, (g_{ab} + u_a u_b)\,,
\end{equation}
where the dots stand for $t$ derivatives and $\nabla$ is the covariant derivative operator in the constant $\psi$ slices.
The energy density $\rho$ can be obtained directly from the Friedmann equation:
\begin{equation}\label{Friedmann_general}
   \left( \frac{\dot{R}}{NR}\right)^{2} = \frac{\rho}{3}-\frac{k}{R^{2}}\,.
\end{equation}

In the evolution formalism, the vacuum field equations can be written as~\cite{4+1}:
\begin{itemize}
  \item The \textit{scalar constraint}:
  \begin{equation}\label{scalar constraint}
    {K_a}^b{K_b}^a - (trK)^2 =\, ^{(4)}\!R\,.
  \end{equation}
  \item The \textit{vector constraint}:
  \begin{equation}\label{vector constraint}
    \nabla_b\,[\,{K_a}^b - trK \,{\delta_a}^b\,] = 0\,.
  \end{equation}
  \item The \textit{evolution equations}:
  \begin{equation}\label{evolution eqs.}
    \partial_\psi~{K_a}^b = \epsilon\,\nabla_a\,\partial{\,^b}\,A
    + A\;[\,-\epsilon\,^{(4)}\!{R_a}^b + trK \,{K_a}^b\,]\,,
  \end{equation}
  \end{itemize}
where $^{(4)}\!R_{ab}$ is the Ricci tensor of the FRW metric.

Allowing for the degenerate algebraic structure of the extrinsic curvature (\ref{extrinsic curvature}), the scalar constraint amounts to:
\begin{equation}\label{4D scalar curvature}
    ^{(4)}R = \rho-3p = 0\,,
\end{equation}
which implies that the FRW metric obtained in every constant $\psi$ slice is a pure radiation metric.


\section{General solution: The M-metric}

Let us now consider the vector constraint (\ref{vector constraint}). Allowing for (\ref{extrinsic curvature}) and (\ref{uderivatives}), one gets
\begin{equation}\label{Nderiv}
    \dot{R} \; N' = 0\,.
\end{equation}
We will not consider here the trivial case ($R=const$), for which every projection of the bulk metric would lead to Minkowsky space on the brane. We will rather conclude that $N=N(t)$ so that the extrinsic curvature (\ref{extrinsic curvature}) actually vanishes, namely
\begin{equation}\label{Kzero}
    K_{ab} = 0\,.
\end{equation}
The factor $N(t)$ can be eliminated by a suitable redefinition of the $t$ parameter in the 5D metric (\ref{bulk metric}). We get then:
\begin{equation}\label{bulk metric2}
    ds^2 = \epsilon\, A^2(\psi,t)\, d\psi^2 - dt^2 + R^2(t)\;\gamma_{ij}\; dx^i\,dx^j\,.
\end{equation}
The evolution equation (\ref{evolution eqs.}) reads now
\begin{equation}\label{evolution consistency}
    \frac{1}{A}\,\nabla_a\,\partial{\,^b}\,A = ^{(4)}\!{R_a}^b\,,
\end{equation}
where the A derivatives are computed on the constant $\psi$ slices, that is
\begin{equation}\label{Aderivs}
    \partial_a\,A = - \dot{A}\, u_a\,.
\end{equation}

Allowing for (\ref{4D scalar curvature}), the Ricci tensor in the FRW slice corresponds to the pure radiation case:
\begin{equation}\label{4D Ricci}
    ^{(4)}\!{R_a}^b = \frac{C}{R^4}\,({\delta_a}^b+4\,u_au^b)\,,
\end{equation}
where $C$ is an arbitrary constant. According to the Friedmann equation (\ref{Friedmann_general}), the expansion factor verifies
\begin{equation}\label{Friedmann}
    \dot{R}^2 = \frac{C}{R^2} - k\,.
\end{equation}
The space components of (\ref{evolution consistency}) can now be written in a simple form:
\begin{equation}\label{evol space}
    \frac{\dot{A}}{A}\,\dot{R}= -\frac{C}{R^3} = \ddot{R}\,,
\end{equation}
which can be explicitly solved:
\begin{equation}\label{A solution}
    A(\psi,t) = \lambda(\psi)\,\dot{R}.
\end{equation}
The integration factor $\lambda$ can be easily removed by a suitable definition of the variable $\psi$. The remaining components in (\ref{evolution consistency}) provide no additional restriction. As a result, the general vacuum solution for the cosmological case (\ref{bulk metric}) can be written, in explicit form, as:
\begin{equation}\label{M metric}
    ds^2 = \epsilon \left( \frac{C}{R^2} - k\right) d\psi^2 - \left( \frac{C}{R^2} - k\right)^{-1}\!dR^2 + R^2\,\gamma_{ij}\, dx^i\,dx^j.
\end{equation}
We have chosen here the $(\psi,R)$ coordinate pair, because in this way all metric coefficients are fully specified. This shows that the general solution (\ref{M metric}) is actually a single vacuum metric, which we will call 'M-metric' in what follows. A single 'mother' metric in the bulk for the full set of embedded FRW metrics, which can be recovered by projecting (\ref{M metric}) onto different, infinitely-many, 4D hypersurfaces (branes). The particular solution for the $\epsilon = -1$, $k=0$ parameter choice has been recently published~\cite{3+2 paper}.

Let us note that the M-metric (\ref{M metric}) has a Killing vector
\begin{equation}\label{Killing}
    \xi \equiv \partial_\psi\,.
\end{equation}
This implies that the $\psi$-coordinate lines have an intrinsic geometrical meaning. On the other hand, from the physical point of view, the time coordinate is precisely the cosmological expansion factor $R$. This intrinsic meaning, both from the geometrical and the physical point of view, allows an straightforward comparison with other forms of the same metric, like the ones proposed in the pioneering work of Ponce de Leon~\cite{JPL88}.


\section{Non-trivial projections: Regular, Emergent and Signature-changing Universes}

We have seen that the trivial projection on $\psi= const$ hypersurfaces leads to pure radiation FRW universe. Another option is to select instead a nontrivial hypersurface in order to get completely different FRW models.
In order to visualize the behaviour of the resulting 4D projections, we will switch to a new time coordinate, namely
\begin{equation}\label{u coord}
    u = \int_0^R \frac{L^2dL}{C-k\,L^2}\,,
\end{equation}
so that the $(\psi,u)$ sector in the M-metric (\ref{M metric}) gets a explicit conformally-flat form:
\begin{equation}\label{M metric conformal}
    ds^2 = \left( \frac{C}{R^2} - k\right )(\epsilon\,d\psi^2 - du^2) + R^2(u)\,\gamma_{ij}\, dx^i\,dx^j.
\end{equation}

\begin{figure}[h]
\centering
\includegraphics[width=7.8cm,height=5.0cm]{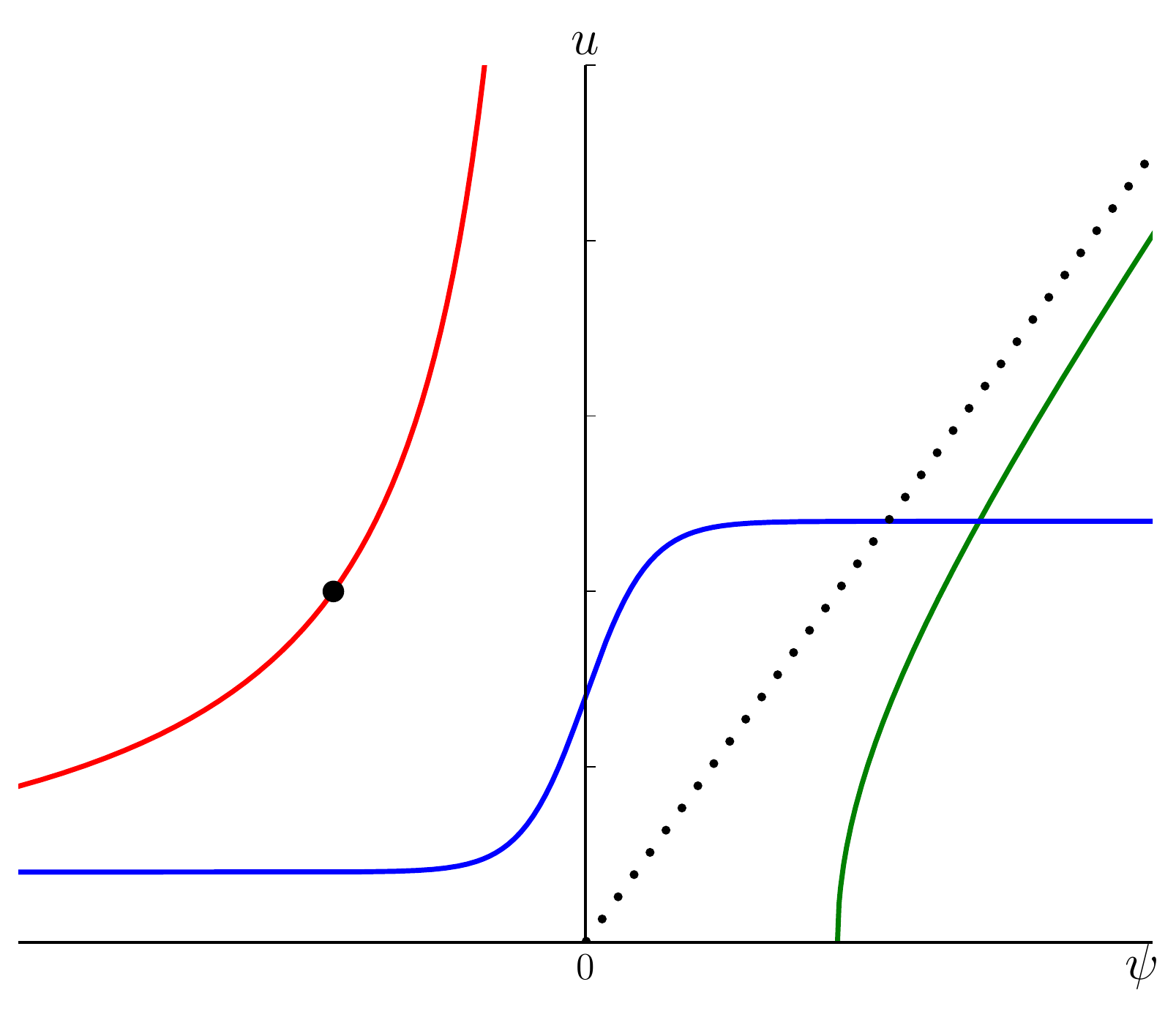}
\caption{ Timelines in the $(\psi,u)$ plane of the bulk M-metric, each one leading to a different FRW brane projection. The big-bang singularity is here the $u=0$ line. The dotted line marks the unit slope, as required from the causality condition arising in the $\epsilon = +1$ case. The green line on the right corresponds to a standard, open big bang model. The blue line in the center corresponds to an emergent universe ($\epsilon = -1$ case). The red line on the left corresponds either to a regular FRW model, without big bang ($\epsilon = -1$) or to a 4D metric showing a smooth transition, at the point marked with a dot, from Euclidean to Lorentzian signature ($\epsilon = +1$).}
\label{timelines}
\end{figure}

We can consider now projections defined by different choices of
\begin{equation}\label{Phi projection}
    \phi(\psi,u) = constant\,,
\end{equation}
which can be visualized as curves in the $(u,\psi)$ plane (see figure~\ref{timelines}).
Note that in the 4+1 case, we must fulfill the additional causality condition
\begin{equation}\label{causality}
|du/d\psi| > 1\;\;\;\;\;\;(\epsilon = +1)\,.
\end{equation}
Otherwise, the projected (brane) metric would be of Euclidean type, instead of the Lorentzian one.

We show in figure~\ref{timelines} some qualitatively different projections. The green line in the right-hand-side corresponds to a quite standard model, defined by
\begin{equation}\label{hyperbola}
u = \sqrt{\psi^2-\phi^2}
\end{equation}
on the constant $\phi$ hypersurfaces. The resulting FRW universe shows a Big Bang singularity ($u=0$) and verifies everywhere the causality condition (\ref{causality}): it will work the for both choices of the signature parameter 

A very different model is obtained by taking instead
\begin{equation}\label{hyperbolic}
    u = \phi + 2\,\tanh{\psi} \,,\;\;\;\;\phi > 2
\end{equation}
(blue line in the center). Note that this choice violates the causality condition everywhere, so it can only work for the $\epsilon=-1$ signature choice. This case corresponds to a regular universe, with no Big Bang singularity. It can be thought as an approximation to some 'Emergent Universe', as defined by Ellis and Maartens~\cite{Ellis_Maartens, Ellis}. The universe expansion starts here from a quasi-stationary state and makes a smooth transition to another quasi-stationary state with a higher value of the expansion factor.

Finally, the red line in the left-hand-side, given by
\begin{equation}\label{regular}
    u = -\phi^2/\psi
\end{equation}
corresponds also to a regular universe, but it verifies the causality condition only in the upper part of the plot ($u>\phi $). This is not a problem in the $\epsilon=-1$ case ($\psi$ timelike): there is no beginning, as the big bang singularity is just in the infinite past limit. Although the singularity is not actually reached, the physical conditions near there can be very close to those just after the big bang in standard models.

In the $\epsilon=+1$ case ($\psi$ spacelike), however, we get a completely different description. There is a dynamical signature change: a smooth transition from an Euclidian to a Lorentzian 4D metric in the brane. This change of signature can be considered~\cite{Ellis_signature,Mars} a classical counterpart of the Hartle-Hawking 'no-boundary' proposal~\cite{Hartle_Hawking}, without resorting to the 'imaginary time' idea. The resulting Lorentzian (FRW)  model will then show an initial singularity (both density and the Hubble factor blow-up) at a finite value of the expansion parameter ($u=\phi$).


\section{Matter-Energy content on the brane: realistic equations of state}

Let us consider now the general FRW line element. Using the scale factor $R$ as the time coordinate, and allowing for the general Friedmann equation (\ref{Friedmann_general}), we get
\begin{equation}\label{FRW metric}
    ds^2 = -\left( \frac{\rho}{3}\,R^2 - k\right)^{-1}\; dR^2 + R^2\;\gamma_{ij}\; dx^i\,dx^j\,,
\end{equation}
where $\rho(R)$ is the density, and the pressure $p$ is given by
\begin{equation}\label{pressure}
    \frac{d\rho}{dR} + 3\,\frac{\rho + p}{R}= 0\,.
\end{equation}
We can see that the explicit form of the FRW metric is provided just by specifying the total density on the brane in terms of the scale factor. Allowing for (\ref{pressure}), this can be done by imposing any suitable equation of state. In the barotropic case $(p=w\rho)$ we get then:
\begin{equation}\label{barotropic}
    \rho \propto R^{-3(1+w)}\,,
\end{equation}
where $w$ is the barotropic index.

We can compare this form of FRW metrics with the projection of the bulk metric (\ref{M metric}) onto an arbitrary (brane) hypersurface, given by (\ref{Phi projection}). We get:
\begin{equation}\label{M metric projection}
    ds^2 = \left[ \epsilon (\frac{C}{R^2} - k)(\frac{d\psi}{dR})^2 - (\frac{C}{R^2} - k)^{-1}\right] dR^2 + R^2\,\gamma_{ij}\, dx^i\,dx^j.
\end{equation}
A simple calculation shows that the two metrics, (\ref{FRW metric}) and (\ref{M metric projection}), match if and only if $d\psi/dR$ verifies
\begin{equation}\label{matching condition}
    \epsilon\left[ (\frac{C}{R^2} - k)\, (\frac{d\psi}{dR})\right]^2 = \frac{\rho\,R^4-3\,C}{\rho\,R^4-3\,k R^2}\,.
\end{equation}

At first sight, it seems that (\ref{matching condition}) guarantees that any FRW metric can be embedded in the cosmological 5D bulk (\ref{M metric}). But one must pay attention to the signature restrictions contained in (\ref{matching condition}), namely
\begin{equation}\label{signature restriction}
    \epsilon = \sgn(\,\rho\,R^4-3\,C\,)\,.
\end{equation}

\begin{table}[]
\begin{tabular}{||c||c|c|}
\hline
Signature                                                          & $\epsilon=1$                   & $\epsilon=-1$                 \\ \hline
\begin{tabular}[c]{@{}l@{}}\quad$R\to 0$\\(Big Bang)\end{tabular}      & $\omega_{\max}\geq \frac{1}{3}$ & $\omega_{\max}\leq\frac{1}{3}$ \\ \hline
\begin{tabular}[c]{@{}l@{}}\qquad$R\to\infty$\\(Open Universe)\end{tabular} & $\omega_{\min}\leq \frac{1}{3}$ & $\omega_{\min}\geq \frac{1}{3}$                              \\ \hline
\end{tabular}
\caption{Restrictions on the adiabatic index range from different asymptotic regimes, for every signature choice.}
\end{table}

The consequences of this restriction for a generic mixture of adiabatic fluid components with adiabatic indexes in the range $\omega_{min} \leq \omega \leq \omega_{max}$ are displayed in Table I. In the 4+1 case ($\epsilon = -1$), there is no problem for a generic equilibrium combination of matter, radiation and cosmological constant terms (with positive density values), namely
\begin{equation}\label{density}
    \rho = \rho_{matter} + \rho_{rad} + \Lambda\,.
\end{equation}
But note that the restriction on $\omega_{max}$ implies that the presence of a radiation component (or something stiffer) is actually required. For the mixture (\ref{density}) this means
\begin{equation}\label{C choice}
    3\,C < \rho_{rad}\,R^4\,.
\end{equation}
The specific expression for $\psi(R)$ can be obtained then from (\ref{matching condition}), modulo a quadrature.

In the $\epsilon=-1$ signature case (\,3+2 bulk manifold), we find just the opposite situation, as the restrictions coming from the two asymptotic regimes are incompatible (except in the pure radiation case). Reasonable physical models (with something more than radiation components) will require then a closed universe ($k=+1$). In this case the only restriction will be $\omega_{max}\leq\frac{1}{3}$, which
would forbid components stiffer than radiation in the Big Bang regime.

This limitations go against the common belief that the Campbell theorem~\cite{Campbell} (see ref.~\cite{Seahra} for its extension to the pseudo-Riemannian case) ensures the embedding of any four-dimensional metric into a five-dimensional Ricci-flat manifold, where the extra dimension can be either space-like or time-like. Let us stress here that what the theorem really says is that any 4D metric can be embedded in a 5D Ricci-flat manifold \textit{provided that} the embedding equations (\ref{scalar constraint}-\ref{evolution eqs.}) hold at least on a single 4D hypersurface. Our results show that this assumption is actually not fulfilled for the density choice (\ref{density}) with the signature combination $\epsilon=-1$, $k=0,-1$.

\subsection*{A 'Big unfreeze' simple model}
We will here provide a simple example of a FRW model showing a dynamical signature change. It can be obtained from the projected metric (\ref{M metric projection}), although we will consider here for simplicity just the spatially flat ($k=0$) case, namely:
\begin{equation}\label{Friedmann projection}
    ds^2 = -C\left[\frac{R^4}{C^2} -(\frac{d\psi}{dR})^2 \right]\frac{dR^2}{R^2}  + R^2\,\gamma_{ij}\, dx^i\,dx^j\,,
\end{equation}
where the energy density can be obtained by direct comparison with the standard FRW expression (\ref{FRW metric}), that is
\begin{equation}\label{model density}
    \frac{3}{\rho} = C\left[\frac{R^4}{C^2} -(\frac{d\psi}{dR})^2 \right]\,.
\end{equation}

The trivial $\psi=constant$ choice leads to a standard pure-radiation model, that is
\begin{equation}\label{radiation}
    \rho = \frac{3C}{R^4}\,.
\end{equation}
A related 'Big unfreeze' model could be for instance
\begin{equation}\label{radiation unfreeze}
    \rho = \frac{3C}{R^4-R^4_0}\,,
\end{equation}
which can be obtained from the simple linear relation
\begin{equation}\label{linear recipe}
    \frac{d\psi}{dR} = \pm \frac{R_0^2}{C}\,.
\end{equation}

It is clear here that a signature change will occur when $R=R_0$. At this point, the energy density becomes infinite, but it just marks the transition from the Riemannian to the pseudo-Riemannian signature in the FRW projected metric (\ref{Friedmann projection}): the beginning of time. Of course, more sophisticated models can be obtained from different generalizations of the linear prescription (\ref{linear recipe}). For instance, one could get easily
\begin{equation}\label{simple density}
    \rho = \frac{3C + \Omega R + \Lambda\,R^4}{R^4-R^4_0}\,.
\end{equation}
Note that in this case the asymptotic expansion regime $R \gg R_0$ is
\begin{equation}\label{Cosmological constant}
    \rho \simeq \frac{3C + \Lambda\,R_0^4}{R^4}+\frac{\Omega}{R^3}+\Lambda+ O(R^{-7})\,,
\end{equation}
so it can fit the present values of the density parameters.


\section{Conclusions and outlook}

We have found the general solution for metrics verifying the natural extension of the Cosmological Principle to a five-dimensional (bulk) manifold. Apart from the trivial case, there is a unique solution for every curvature sign ($k=0,\pm 1$) and every signature of the extra  coordinate ($\epsilon = \pm 1$): the M-metric (\ref{M metric}). This metric has a Killing vector (\ref{Killing}), and we have used adapted orthogonal coordinates $(\psi,R)$, where $\psi$ is the corresponding cyclic coordinate and $R$ is the cosmological expansion factor. In this way, our coordinates have a sound geometrical and physical meaning.

We have considered the projections of the M-metric onto space-homogeneous four-dimensional hypersurfaces (branes) in order to recover FRW metrics. The use of $R$ as the time coordinate in the brane allows one to express the generic FRW metric directly in terms of the matter-energy density (\ref{FRW metric}). This is crucial for controlling the equation of state of the resulting projection. We have studied the case of a generic (equilibrium) mixture of matter, radiation and cosmological constant. We have provided, modulo a quadrature, the transformation $\psi(R)$ allowing to recover the corresponding metric for both signature combinations ($\epsilon=\pm 1$) in the closed universe case ($k=+1$). In the open case ($k=0,-1$), we have shown that, apart from the pure radiation case, the generic combination can be obtained only when the extra coordinate $\psi$ is space-like.

The $(\psi,R)$ plane in the bulk happens to be a powerful tool for devising the evolution properties of the resulting projected spacetimes: one only has to select a suitable time curve, which will be kept as the physical time coordinate in the brane. We shown how to obtain by this method some FRW regular solutions, evolving from the infinite past (no Big Bang), that could be useful to deal with the Cosmological Horizon problem. These are regular FRW models in standard General Relativity: there is no need to recur to alternative theories in order to get these appealing cosmological models. We have also seen how to obtain in this way models that approach some instances of the well-known 'Emergent Universe' inflationary models~\cite{Ellis_Maartens, Ellis}.

In the 4+1 case, where the extra coordinate is spacelike, we have identified the causality condition (\ref{causality}) which ensures that the projected (brane) hypersurface is of Lorentzian type. When violated, the projection leads to an Euclidean 4D manifold, with no time dimension. We can take advantage of this in order to get universe models showing a dynamical signature change: a smooth transition from an Euclidian to a Lorentzian 4D metric. This change of signature was proposed by Hartle and Hawking in the quantum Cosmology context in order to avoid imposing any initial boundary condition: the wave function could then oscillate in the Lorentzian domain and get exponentially damped in the Euclidean one. In the purely classical context, it has been suggested that the bulk-brane paradigm could lead to this signature change in a natural way: a perfectly smooth solution in the bulk leading to a 4D signature-changing metric in the brane projection~\cite{Ellis_signature,Mars}. The bulk solution, however, was not clearly identified: a bottom-up approach was used, assuming the compatibility of the Euclidean and the Lorentzian parts embedding. We have rather used a top-down approach, starting from the bulk M-metric (\ref{M metric}), and providing explicit expressions for the projection process. The resulting FRW model shows an initial singularity (density blowing-up) at a finite value of the scale factor $R$. This has been termed as 'Big freeze' singularity type~\cite{Jambrina sing} because it has been associated with the final stage of cosmological evolution, as an alternative to the 'Big rip'~\cite{Bouhmadi}. This term is misleading in our case, where we could rather use the term 'Big unfreeze' for this singularity, as it is produced just by the beginning of time, without affecting space geometry. We have shown an explicit example, in the spatially flat case, arising from a simple linear prescription for $\psi(R)$, which can be easily altered in order to get more sophisticated models, that can fit the present values of the density parameters.

As a final remark, let us point out that having the cosmological bulk metric (\ref{M metric}) in fully explicit form can open the door to new advances in some related fields. For instance, a semiclassical Kaluza-Klein approach to Cosmology would clearly benefit from the uniqueness of the spacetime background: any result obtained from the vacuum line element (\ref{M metric}) would allow to draw general conclusions, valid beyond any particular case. This can be important in the study of perturbations that can lead to structure formation. In a 4+1 context, the inhomogeneous perturbation could in principle propagate through the extra space dimension, affecting to the power spectrum, which will depart from the one derived from the standard 3+1 calculations. The usual recipe to deal with that problem is to consider the extra space direction to be compact. A simple look to Fig.~1 shows, however, that the time lines are not periodic, so one can not assume that the $\Psi$ coordinate is compact. But note that the coordinate freedom in the bulk allows one to move from the $(\Psi,R)$ coordinates (R defining the ‘expansion time’) to some other $(\Phi,t)$ coordinates, where t will be the coordinate adapted to the selected ‘physical’ timeline, defined then by $\Phi(\Psi,R)= constant$. The natural choice will be to compactify precisely this $\Phi$ space coordinate, orthogonal in the bulk to the physical timelines. In this way, the choice of the special (compact) space coordinate amounts to the choice of the corresponding (orthogonal) timeline in the bulk which will lead to a specific FRW model in the brane. For more potential applications, see for instance ref.~\cite{Wesson_book}.


\acknowledgments
We acknowledge support from the Spanish Ministry of Economy, Industry and Competitiveness grants AYA2016-80289-P and AYA2017-82089-ERC (AEI/FEDER, EU). MB would like to thank CONICYT Becas Chile (Concurso Becas de Doctorado en el Extranjero) for financial support.




%
%

\bibliographystyle{prsty}

\end{document}